\documentclass[a4paper,11pt]{article}
\pdfoutput=1 

\usepackage{jheppub} 

\usepackage[T1]{fontenc} 

\title{Quantization of horizon entropy and the thermodynamics of spacetime}

\author[]{Jozef Sk\'akala}

\affiliation[]{UFABC, Santo Andr\'e, Sa\~o Paulo, Brazil}

\emailAdd{jozef.skakala@ufabc.edu.br}

\abstract{This is a review of my work published in the papers  \cite{Skakala2, Skakala3, Skakala5, Skakala4}.  It offers a more detailed discussion of the results than what was given in the published papers and it links my results to some conclusions recently made by other people. It also offers some new arguments for the conclusions previously made. The fundamental idea of this work is that the semi-classical quantization of the black hole entropy, as suggested by Bekenstein \cite{Bekenstein1}, holds (at least) generically for the spacetime horizons. We support this conclusion by two separate arguments: ~1. we generalize Bekenstein's lower bound on the horizon area transition to much larger class of horizons than only the black hole horizon, ~2. we obtain the same entropy spectra via the asymptotic quasi-normal frequencies of some particular spherically symmetric multi-horizon spacetimes, (in the way proposed by Maggiore \cite{Maggiore}). The main result of this paper supports the conclusions made by other people in \cite{Paddy3, Kwon}, but uses different arguments.}

\begin{document}

\maketitle
\flushbottom

\section{Introduction}

This paper is a review of some of the work I have done during my postdoctorial fellowship in Brazil. It summarizes the basic results from some of the papers I published during the given period \cite{Skakala2, Skakala3, Skakala5, Skakala4}. It gives a more detailed discussion of the results than what was given in the published papers and it connects the results from the papers \cite{Skakala2, Skakala3, Skakala5, Skakala4} to some conclusions recently made by other people. It also presents some new results, such that provide additional support for the basic idea presented in this work. The fundamental idea of this work is that the semi-classical quantization for black hole entropy as suggested by Bekenstein \cite{Bekenstein1} holds (at least) generically for the spacetime horizons and that the asymptotic quasi-normal frequencies carry information about the quantized spectra (in the way suggested by Maggiore \cite{Maggiore}). 

The structure of this paper goes as follows: first we offer a brief introduction to some of the main results from the area of spacetime thermodynamics. Then we give some arguments in support of the connection between the asymptotic quasi-normal frequencies and the quantization of entropy, as suggested by Maggiore \cite{Maggiore}. (Maggiore refined the original conjecture of Hod \cite{Hod}.) In the following section we show that the lower bound on the area / entropy increase derived by Bekenstein in his seminal work \cite{Bekenstein1} can be applied to a much larger class of spacetime horizons. These results are followed by the analysis of the behavior of asymptotic quasi-normal frequencies of some of the spherically symmetric multi-horizon spacetimes and interpreting the behavior along the way suggested by Maggiore. In all the sections we offer a relatively detailed discussion of our results. 

\subsection{The brief overview of the spacetime thermodynamics}

The first origins of the spacetime thermodynamics trace down to the
period more than 40 years ago, to the formulation of black hole no
hair conjecture and the discovery of Penrose
process \cite{Penrose}, which shows that certain number of energy to be extracted from
a rotating black hole. The fact that the black hole energy can decrease came as a surpirse, and the natural question that came up after
Penrose discovery was, whether there are, and what are the parameters of the isolated black hole that cannot
decrease in any classical physical process. Answer to this question
came with the work of \cite{Hawking, Christodulou, ChristodulouRuffini}, which showed the black hole horizon area (or, consequently, the irreducible mass),  cannot
decrease during the classically allowed physical processes.

The result of \cite{Hawking} was subsequently followed by a very
interesting insight of Bekenstein \cite{Bekenstein1}, suggesting that black hole
horizon area behaves in the same way as one would expect from the black hole entropy
(proposing their proportionality, such that a preferable option is linear dependence $S\sim A$). The Bekenstein's
suggestion corresponded to the four laws of black hole mechanics
derived in the work of \cite{fourlaws}. (Let us mention that the nature of what provides the degrees freedom for the black hole
entropy is an intense ongoing debate since.) The four laws of black
hole mechanics provided a complete analogy with the classical four
laws of thermodynamics. They suggested, if taken seriously, that
black hole has both, the entropy proportional to the area of its
horizon, (as proposed by Bekenstein), and a temperature proportional
to the black hole surface gravity.  The four laws of black hole
mechanics were first seen just as an curious analogy, until Hawking
showed that the analogy is in fact identity by proving that black
holes semi-classically radiate \cite{Hawking2, Hawking3}. Hawking's result fixed the
proportionality of the black hole temperature and the surface
gravity to be (in Planck units) $T=\kappa/2\pi$, and the entropy
with the horizon area to be $S=A/4+const.$~ (Note that everywhere in what
follows we will use the Planck units. This will be
reminded few more times on few different places in the text.)

Since the early years, the field of spacetime thermodynamics
conceptually developed in various directions. First, a
generalized second law of black hole thermodynamics was proposed
\cite{Bekenstein2}. Second, it has been shown by different authors that not only the black hole horizon, but generally
spacetime horizons radiate with a temperature proportional
to their surface gravity, (as related to a suitably normalized Killing
field). This was for example shown for the accelerated observers
with the Rindler horizons \cite{Unruh}, for the cosmological horizon \cite{HawkingGibbons}, and recently it was suggested that even Cauchy horizons
could radiate \cite{Makela}. This means the quantum
radiation is a characteristic feature of spacetime horizons in
general, (independent on the fact whether the horizons are
\emph{absolute}, or \emph{apparent} and independent on the general relativity theory \cite{Paddy1}), rather than being exclusively
a property of the black hole horizon. (For general results related
to static spherically symmetric spacetimes see for example \cite{Paddy1}) 

This way of
thinking was further extended in the works of \cite{Jacobson, Paddy1, Paddy2, Paddy4} which suggest, that not only the concept
of temperature, but also the concept of horizon entropy\footnote{Let us also mention here that the generalization of the concept of black hole entropy to arbitrary gravity theory given by a diffeomorphism invariant Lagrangian was given by Wald \cite{Wald}.} can be
understood more generally as a property of spacetime horizons. (This
is again independent on whether the horizon is observer dependent,
or not.) This simply means, that it is more appropriate to speak about
``spacetime thermodynamics'', rather than exclusively about ``black
hole thermodynamics''. A lot of support to such ideas is
given by the fact that the first law of thermodynamics can be
formulated reasonably generally,
by using only quasi-local concepts \cite{Paddy1, Paddy2, Jacobson}, (without assumptions such as asymptotic flatness). For example Padmanabhan \cite{Paddy2} had shown, that Einstein equations when evaluated at
a horizon in a static spherically symmetric spacetime, reduce to a thermodynamical
identity. (This can be generalized to stationary axisymmetric spacetimes \cite{Paddy1}.)

The interesting question to be answered was: How does the classical
general relativity know about the thermodynamics of spacetime? The
interesting answer to this question was suggested in
\cite{Jacobson}: The general theory of relativity could be just a continuum
statistical description of fundamentally different degrees of
freedom. Just like sound waves in a medium statistically relate (in a
thermodynamic equilibrium) to the atoms and molecules, those being the real
degrees of freedom on the more fundamental level. Taking such a
viewpoint, it is not so surprising that general relativity
inherently contains some information about parameters related to the equation of
state of those degrees of freedom.

\subsection{Bekenstein's semi-classical quantization of horizon entropy}

One can try to ``dig out'' even more subtle information about the
thermodynamical variables. This relates to the third important
direction in which the early ideas evolved: The Bekenstein's
ideas about the semi-classical quantization of the black hole
horizon area (and consequently black hole entropy). Bekenstein in
his early work \cite{Bekenstein1} noticed, that Christodulou's reversible process, describing a
particle being radially dropped under the horizon, such that the
particle's classical turning point is put infinitesimally close
to the horizon, leads to a conclusion that black hole horizon area
behaves as a classical adiabatic invariant. One can then use the Ehrenfest
principle from the early days of quantum mechanics. The Ehrenfest principle states
that the classical adiabatic invariants should have semi-classically
discrete quantum spectra, and therefore Bekenstein proposed that the black hole
horizon area is supposed to have a semi-classically discrete
spectrum. 

Furthermore, by considering a quantum modification of the
classical reversible Christodulou process, Bekenstein concluded that
there is in fact a non-zero lower bound on the horizon area
increase. This means, placing the particle's
classical turning point infinitesimally close to the horizon
contradicts the quantum principle of uncertainty, and the reversible
processes are not allowed. Bekenstein obtained the lower bound for
the horizon area increase in Planck units as $\Delta
A_{min}=8\pi$. The universality of this bound (independence on the black hole parameters) led Bekenstein \cite{Bekenstein1} to a conclusion that the preferred option is the \emph{equispaced} area spectrum. The quantum black hole horizon area
spectrum can be simply conjectured from: ~1. the existence of a universal, black hole parameter independent lower
bound on the area change, such that the minimum area change can be
in certain limiting processes always approximated, ~2. the spectrum is reasonably simple and therefore the area quanta in the spectrum do not form an
irregular, or oscillatory sequence (sequence of numbers approximately around the value of the bound). Considering these two requirements
on the area spectrum and the Bekenstein's bound, one uniquely obtains Bekenstein's \cite{Bekenstein1, Bekenstein3} semi-classically
equispaced spectrum given in Planck units as
\begin{equation}\label{Aspectrum}
A_{n}=8\pi\gamma\cdot n, ~~~~~~~~~~~~\gamma\in O(1).
\end{equation}
(The uncertainty in the value of $\gamma$ represents some uncertainties in the derivation of the lower bound.)

The suggestion given by \eqref{Aspectrum} means that one obtains (also) the entropy of
the black hole horizon semi-classically quantized and
equispaced. But claiming that such an equispaced spectrum of entropy
should be trusted beyond the semi-classical approximation, would
have significant consequences: the statistical interpretation of entropy, (logarithm of the microscopic degeneracy of a
macroscopic state), means the quantum of entropy must be of the form
$\Delta S=\ln(k), ~~~k\in\mathbb{N_{+}}$~ \cite{BekensteinMukhanov}. On the other hand, over the years different black hole
quantization schemes appeared (see for example \cite{Maggiore0, Makela2, Kunstatter1, Kunstatter2, Medved2}), most of them in favor of
the equispaced area / entropy spectrum, but largely agreeing on
fixing the $\gamma$ value in \eqref{Aspectrum} as $\gamma=1$. (This would mean the bound was fixed in some sense precisely.) Such
results might indicate that one should be probably less ambitious, and
keep the suggestions for the equispaced entropy spectra only within
the realms of the semi-classical approximation \cite{Maggiore}.

One could then naturally ask if the Bekenstein semi-classical
quantization scheme is only a property of the black hole horizons, or a
more general property of the spacetime horizons. And, indeed, even the Bekenstein semi-classical entropy
quantization law is suggested to apply generally to spacetime horizons \cite{Paddy3}. It had been shown in \cite{Paddy3} that for a
general Lanczos-Lovelock gravity theory, one can impose certain reasonable
physical assumptions, and obtain an equispaced semi-classical entropy spectrum for
general spacetime horizons. The spectrum is for every horizon of
what seems to be the most popular form: $S_{n}=2\pi\cdot n$, (in
Planck units). Outside the realm of general relativity, equispaced entropy spectrum does
not necessarily give the equispaced horizon area spectra, but
considering only Einstein's gravity, the quantization of the
spacetime horizon area becomes $A_{n}=8\pi\cdot n$. 

The reasoning
(for more details see \cite{Paddy3}) goes as follows: Any effective action of the
observer constrained to the accessible region must depend only on
the information accessible to that observer. Then consider any kind
of spacetime horizon (horizon, as perceived by the observers linked
to the coordinates singular at the given horizon). The requirement
that the observer constrained by such a horizon must require only
information accessible to the observer, leads in the semi-classical
WKB approximation to the fact, that an effective action of such an
observer contains a surface term, which evaluated on the horizon
encodes all the information hidden to the observer. Such a boundary term is not generally covariant and should not have any effect on the quantum effects described by the observer. For the boundary term to fulfill this condition it has to be semi-classically quantized as:
\[A_{sur}=2\pi n,~~~~~~n\in\mathbb{Z}.\]
In Einstein's theory this term is just the entropy of the horizon and therefore gives the uniform quantization of entropy with the quantum $\Delta S=2\pi$. (But this entropy spectrum turns out to be general within the Lanczos-Lovelock gravity theories \cite{Paddy3}.)

\subsection{The conjectured connection to the asymptotic quasi-normal modes}

Additional evidence for the equispaced horizon area / entropy
spectra appeared around 15 years ago in the works of Bekenstein \cite{Bekenstein4} and
Hod \cite{Hod}. Take some specific field (scalar, electromagnetic,
gravitational etc.) and let it scatter (for example) on the Schwarzschild black
hole spacetime. The scattering amplitude has a discrete, infinite
number of non-real complex poles, the quasi-normal frequencies, such
that are labeled by the wave mode number $\ell$ and an overtone
number $n$. (The frequencies are most conveniently computed by
decomposing the perturbation in the tensor spherical harmonics.) Some
combination of the frequencies dominates the time evolution of an
arbitrary perturbation (coming from compactly supported initial
data) within a characteristic time scale. This is the reason, why
the quasi-normal frequencies are often called the ``ring-down'' frequencies,
or ``the characteristic sound of black holes''. (For more details of what are the quasinormal modes see for example the reviews \cite{Nollert, Cardoso, Zhidenko}.)

If one fixes  $\ell$, the quasi-normal frequencies still form an
infinite discrete set, with an unbounded imaginary part (they
describe arbitrarily highly damped oscillations). Let us focus on the asymptotic sequence of the frequencies with the
damping going to infinity, (this is what we further call \emph{asymptotic} frequencies). These frequencies show for the Schwarzschild black hole, perturbation with
spin $s$ and the wave mode number $\ell$ the following behavior \cite{Motl, Andersson, Zhidenko}:
\begin{equation}
\omega_{s,n,\ell}=\hbox{(offset)}_{s}+i\cdot 2\pi T\cdot
n+O(n^{-1/2}).
\end{equation}
By $T$ we mean here the Hawking temperature of the Schwarzschild
black hole horizon. This is a remarkable universal feature:
The spacing of the asymptotic sequence of the modes is completely
independent on the type of perturbation, or the wave mode number. (Note that this is in fact a general feature of spherically symmetric, static asymptotically flat spacetimes, not even constrained by the general relativity \cite{Shanki1, Medved}.)
Furthermore, the ``$\hbox{(offset)}$'' is some complex number, such that it is also
independent on the wave mode number and depends only on the spin of
the perturbation. The fact that the spacing between the asymptotic
frequencies is universally converging to a constant given as ~$2\pi
T$~ suggests the link of the frequencies to the Euclidean gravity \cite{Motl}. (The
Euclidean metric is periodic with the period given by the inverse
temperature.) 

It was suggested in the works
of Bekenstein \cite{Bekenstein4} and Hod \cite{Hod}, that by using Bohr's correspondence principle: ``the transition frequencies at high quantum numbers equate to the classical oscillation frequencies'',
the asymptotically highly damped modes could link to the quantum black hole properties (in the semi-classical regime). In
particular Bohr's correspondence principle links the asymptotic quasi-normal frequencies to the
quanta of energies emitted in the black hole state transitions. (A more detailed discussion we give later in the paper in the section \ref{discussionMaggiore}.)
However the original Hod's conjecture \cite{Hod} was later refined by a suggestion of
Maggiore \cite{Maggiore}. Maggiore proposed \cite{Maggiore} to treat the asymptotic quasi-normal
frequencies as a collection of damped oscillators with proper
frequencies given as, (let us further use the notation $\omega=\omega_{R}+i\omega_{I}$):
\[\sqrt{\omega_{nR}^{2}+\omega_{nI}^{2}}.\]
(As mentioned before, we give a more detailed reasoning why this connection might be relevant in the section \ref{discussionMaggiore}. For some further discussion we recommend also the original papers \cite{Hod, Maggiore}. Also for some alternative, but complementary viewpoint about the connection of the asymptotic modes to the entropy/area spectra see \cite{KunstatterQNM}.) 
As we will see, Maggiore's link of the hypothetical
quanta of energy emitted, or absorbed in a black hole state
transition to the asymptotic quasinormal modes reads as:
\[\Delta M=\lim_{n\to\infty}\Delta_{(n,n-1)}\sqrt{\omega_{nR}^{2}+\omega_{nI}^{2}}.\]
For the Schwarzschild black hole this turns into:
\begin{equation}\label{Mag}
\Delta M=\lim_{n\to\infty}\Delta_{(n,n-1)}\sqrt{\omega_{nR}^{2}+\omega_{nI}^{2}}=\lim_{n\to\infty}\Delta_{(n,n-1)}\omega_{nI}=2\pi T.
\end{equation}

Take
the first law of black hole mechanics and consider the Bekenstein's
suggestion for the quantum of entropy, $\Delta S=2\pi\gamma$. Then
the mass quantum corresponding to the entropy change is given as:
\begin{equation}\label{firstsimple}
T\Delta S=2\pi\gamma\cdot T=\Delta M.
\end{equation}
The equation \eqref{firstsimple} together with Maggiore's conjecture\footnote{This can be called modified Hod's conjecture, but we will keep it for simplicity under the name ``Maggiore's conjecture''.} suggest that the
asymptotic spacing between the quasinormal frequencies of the
Schwarzschild black hole should be given as $2\pi\gamma T$. This
is, as we have seen, precisely the case, if one fixes $\gamma$ in the most popular way: $\gamma=1$. Hence
Maggiore's suggestion gives the same, universal result for the area quantum as the Bekenstein's lower bound on the area transition. This can be seen as a strong additional support for the black
hole area spectrum of the form $8\pi\cdot n$, (entropy quantization
$2\pi\cdot n$). In the same time it explains the ``$2\pi T$'' universality in the
spacing of the asymptotic frequencies. (It matches those two seemingly unrelated results.)

\subsection{The results presented in this paper}

In this paper we are particularly interested in the spacetimes with
multiple horizons, but we keep things sufficiently simple and will
deal only with the spherically symmetric static spacetimes. This work is
supposed to result in support of three basic ideas that were
presented in this introductory section: ~1. the entropy spectra of
the spacetime horizons are all quantized with a quantum given as
$\Delta S=2\pi$, ~2. the asymptotic quasi-normal modes play the role
in determining the spectra, as suggested by Maggiore, ~3. most
importantly, there is no fundamental difference between the
spacetime horizons in terms of their thermodynamical properties. The first
statement just confirms the result of \cite{Paddy3}. We further discuss
some of the consequences of such a universal entropy quantization.
The second point means we extend the use of Maggiore's conjecture to
the multi-horizon case (in an analogy to the Schwarzschild case),
giving an additional support for the conjecture. Via Maggiore's conjecture we explain the observed
behavior of the quasi-normal frequencies in the multi-horizon cases,
and in the same time we argue in favor of the entropy spectra
proposed by \cite{Paddy3}. The third statement is a very natural result,
strongly suggested by the work of different authors \cite{Jacobson, Paddy1}. The fact that
certain thermodynamical ideas appeared first in the context of the
black hole horizons is therefore more a coincidence, rather then
something that reflects any fundamental features of the nature.

\section{The thermodynamical concepts in static spherically symmetric spacetimes}
In this section we will, for the convenience of the reader, briefly introduce the basic equations and results that are of key importance for our work. (The results are standard and can be found over the literature. The reader well familiar with the concepts can skip most of the details.) 

Take a specific type of static spherically symmetric line
element in the fixed ``suitable'' coordinates:
\begin{equation} \label{line}
ds^{2}=-f(r)dt^{2}+f^{-1}(r) dr^{2}+r^{2}d\Omega^{2},
\end{equation}
in some region where $f(r)>0$. (Assume also that $f(r)$ is a well
behaved smooth function in this region.) Let us take in this region
a spacelike hypersurface of the fixed time~ $\Sigma_{t}\doteq
(r,\theta,\phi)$ and consider the Killing field
$K\doteq\partial_{t}$. The Killing field
fulfils the following relation:
\begin{equation}
 K^{a;b}_{~~;b}=R^{ab}K_{b}.
\end{equation}
Then applying Stoke's theorem one obtains:
\begin{equation}
\int_{\Sigma_{t}}K^{a;b}_{~~;b}d\Sigma_{a}=\int_{\partial \Sigma_{t}}K^{a;b}d\Sigma_{a b}.
\end{equation}
One can use then the Einstein equations to conclude the following:
\begin{eqnarray}~\label{Komar}
 \int_{\partial \Sigma_{t}}K^{a;b}d\Sigma_{a b}=\int_{\Sigma_{t}}R^{a b}K_{b}d\Sigma_{a}=~~~~~~~~~~~~~~~~~~~~~~~~~~~~~\nonumber\\
=8\pi\int_{\Sigma_{t}}\left(T^{a b}-\frac{1}{2}Tg^{a b}\right)K_{b}d\Sigma_{a}+\Lambda\int_{\Sigma_{t}}g^{a b}K_{b}d\Sigma_{a}.~~~~
\end{eqnarray}
Choose the boundary $\partial \Sigma_{t}$ to be at ~$r=R_{1}, R_{2}$~ with
$R_{1}<R_{2}$. Then from the equation \eqref{Komar} we can deduce:
\begin{equation}\label{Komar1}
-4\pi \tilde M_{1}+4\pi \tilde M_{2}=8\pi\int_{\Sigma_{t}}\left(T^{a
b}-\frac{1}{2}Tg^{a b}\right)K_{b}d\Sigma_{a}+\Lambda\int_{\Sigma_{t}}g^{a
b}K_{b}d\Sigma_{a}.~~~~
\end{equation}
Here $\tilde M_{1}, \tilde M_{2}$ represent masses of the regions inside
the chosen boundaries. These masses correspond to a quasi-local concept
of mass in general relativity, known also as Komar mass. (Or more
generally one speaks about Komar conserved quantities.)

Write the equation \eqref{Komar1} explicitly with the stationary spherically symmetric electromagnetic field:
\begin{eqnarray}\label{Komar2}
\tilde M_{2}=2\int_{\Sigma_{t}}\left(T_{M}^{a b}-\frac{1}{2}T_{M}g^{a
b}\right)K_{b}d\Sigma_{a}+2\int_{\Sigma_{t}}T_{EM}^{a b}K_{b}d\Sigma_{a}+~~~~~~~~~~~~~~\nonumber\\
+2\cdot\frac{\Lambda}{8\pi}\int_{\Sigma_{t}}g^{a
b}K_{b}d\Sigma_{a}+\tilde M_{1}.~~~~~~~~
\end{eqnarray}
Here $T^{ab}_{M}$ is some matter field stress energy tensor and $T^{a b}_{EM}$ is the electromagnetic stress energy tensor. (The trace of the electromagnetic stress tensor is $T_{EM}=0$.)
We clearly see that the explicitly written integrals on the right
side of the equation \eqref{Komar2} represent the contribution to the mass from the region between the
boundaries. 

To have more insight into the equation \eqref{Komar2}, let us partly confirm it by calculating the relevant integrals on both sides.
The integral with the electromagnetic stress energy tensor can be performed explicitly, ($T_{EM}^{tt}=\frac{1}{8\pi}\frac{Q^{2}}{r^{4}}f^{-1}$) as:
\[\tilde M_{EM}=2\int_{\Sigma_{t}}T_{EM}^{a b}K_{b}d\Sigma_{a}=\int_{R_{1}}^{R_{2}}\frac{Q^{2}}{r^{4}}f^{-3/2}\cdot f^{3/2}\cdot r^{2} dr=\int_{R_{1}}^{R_{2}}\frac{Q^{2}}{r^{2}}dr=- Q^{2}\left[\frac{1}{R_{2}}-\frac{1}{R_{1}}\right]. \]
The second integral on
the right side can be also easily computed as:
\begin{eqnarray}
\tilde M_{\Lambda}=(4\pi)^{-1}\Lambda\int_{\Sigma_{t}}g^{a b}K_{b}d\Sigma_{a}=-(4\pi)^{-1}\Lambda\int_{\Sigma_{t}}f^{1/2}\cdot f^{-1/2}r^{2}\sin(\theta)\cdot dr d\theta d\phi=~~~~\nonumber\\
=-(4\pi)^{-1}\Lambda\frac{4\pi}{3}\left(R^{3}_{2}-R^{3}_{1}\right)=-\frac{\Lambda}{3}\left(R^{3}_{2}-R^{3}_{1}\right)=-2\cdot \frac{\Lambda}{8\pi}\cdot[V(R_{2})-V(R_{1})].~~~~~~~~~~
\end{eqnarray}
Here $V(R)=\frac{4\pi}{3}R^{3}$ is a flat space volume contained in a sphere with radius $R$ and ~$-\Lambda/8\pi$ is the energy density of the cosmological constant term.
The mass $\tilde M$ of a region inside the boundary $\partial
\Sigma_{t} : ~r=R$ can be calculated as:
\begin{eqnarray}
\tilde M=(4\pi)^{-1}\int_{\partial \Sigma_{t}}\Gamma^{t}_{rt}\cdot f(R)\cdot R^{2}dS=~~~~~~~~~~~~~~~~~~~~~~~~~~~~~\nonumber\\
=(4\pi)^{-1}\int_{\partial \Sigma_{t}}\frac{1}{2}f^{-1}(R)\cdot f_{,r}\cdot
f(R)\cdot R^{2}dS=\frac{f_{,r}}{2}R^{2}.~~~~~~~~
\end{eqnarray}

Consider Unruh radiation observed by the accelerated observers. The
four-acceleration of the Unruh stationary ``floating'' observes
(world lines $r=const.$) is
\[ a^{b}=u^{c}u^{b}_{;c} ~~\rightarrow ~~ a^{(t,r,\theta,\phi)}=\left(0,\frac{f_{,r}}{2},0,0\right).\]
This means the mass of the given boundary equals
\[\tilde M=\pm f^{1/2}|a|R^{2}=\pm T_{a}\frac{A}{2},\]
where $T_{a}=f^{1/2}|a|/2\pi$ is the Davies-Unruh temperature, redshifted temperature measured by the
Unruh observer floating at the boundary, (moving along the worldline
$r=R$), and $A$ is area of the spherical boundary. The $\pm$ sign
depends on the sign of the $f_{,r}$ function. In the limiting case
of the boundary being a horizon, one obtains from $T_{a}$ the Hawking
temperature $T_{H}$, and this is a temperature of the
radiation measured by a stationary observer with the
zero acceleration (moving along the geodesics). (In case of
asymptotically flat spacetime, $\Lambda=0$, this is a stationary
observer at the infinity, and in case of $\Lambda>0$, the position of
such an floating observer is at a finite radius given by~
$f_{,r}(R)=0$.)

Let us consider that locally around the boundary there is the
Reissner-Nordstr\"om-deSitter (R-N-dS) geometry. In this case the metric function $f$ reads:
\[f(r)=1-\frac{2M}{r}+\frac{Q^{2}}{r^{2}}-\frac{\Lambda}{3}r^{2}.\]
 Then one obtains
the following result:
\begin{equation}\label{Komar3}
\tilde M=M-\frac{Q^{2}}{R}-\frac{\Lambda}{3}R^{3},
\end{equation}
where the special cases of Schwarzschild, de-Sitter,
Schwarzschild-deSitter (S-dS) and Reissner-Nordstr\"om (R-N) spacetimes are
obtained simply by setting $M=0$, $\Lambda=0$ and $Q=0$ as
appropriate.
Also let us derive for the Reissner-Nordstr\"om-deSitter geometry and a horizon with the area $A$:
\begin{eqnarray}\label{areavar}
\delta A=8\pi\cdot R\delta R=-8\pi\cdot R \left[\frac{f_{,M}}{f_{,r}}\delta M+\frac{f_{,\Lambda}}{f_{,r}}\delta\Lambda+\frac{f_{,Q}}{f_{,r}}\delta Q\right]~~~\rightarrow~~~~\nonumber\\
\rightarrow~~~~~~\pm T\delta S=\delta M-\frac{R^{3}}{6}\delta \Lambda+\frac{Q}{R}\delta Q.~~~~~~~~~~~~
\end{eqnarray}
(Here we used the fact that for a boundary, such that it is a spacetime horizon, we can relate the area of the boundary to the concept of entropy.)
 
At the end of this section, let us mention one key result. By varying \eqref{Komar1} and taking the boundaries to be the black hole and the cosmological horizon (for a spacetime where the two horizons are present) one gets the quasi-local form of the first law of thermodynamics as:
\begin{equation}\label{firstlaw}
\int_{\Sigma_{t}}\delta T_{a b}K^{a}d\Sigma^{b}+T_{BH}\delta S_{BH}=-T_{CH}\delta S_{CH}.
\end{equation}
Here index ``BH'' means the black hole horizon and ``CH'' means
the cosmological horizon. The gauge chosen to compare the
solutions is such that $\delta K^{a}=0$.

\section{The generalization of the Bekenstein's lower bound on the horizon area transition}

Let us note that a Killing energy absorbed or emitted by the
horizons can be (in a non-trivial sense) arbitrarily small, which
leads to the conclusion that the horizon area behaves in certain cases
as a classical adiabatic invariant. To clarify this, let us show an
example: A neutral point particle can be described by a following stress
energy tensor
\[T^{a b}=m\cdot \delta\{r-R(t)\}\cdot (u^{0})^{-1} u^{a}u^{b}.\]
($m$ is rest mass of the particle.) Integrating such a distribution
of energy along a hypersurface $\Sigma_{t}$ leads to the following
Killing energy:
\[E=m \sqrt{f(R_{p})},\]
where $R_{p}$ is a classical turning point, at which the point
particle's four-velocity is orthogonal to $\Sigma_{t}$. Closer
the turning point lies to the horizon, smaller is the Killing energy by which
the particle contributes to the area change of the horizon. (As can be seen from the equation \eqref{firstlaw}.)  As a result one can regard the horizon area
as a classical adiabatic invariant. 

This argument was originally put up for the area of the Kerr black
hole horizon \cite{Bekenstein1}, but as we can easily see, it holds much more generally. As an example we can take pure deSitter spacetime and drop
a particle from a turning point infinitesimally close to the
cosmological horizon. Then also the cosmological horizon area would behave as a
classical adiabatic invariant.
Bekenstein used the classical adiabatic invariance and the
Ehrenfest principle to deduce that the black hole horizon area must
have semi-classically discrete quantum spectrum. But then it is reasonable to suggest that discrete spectra should be a general property of spacetime horizons! 

Furthermore, as
already mentioned, Bekenstein argued that going beyond the classical
physics and employing the quantum principle of uncertainty modifies
the classical results. This is because to keep the horizon's area
unchanged one would need to know in the same time the exact particle's
momentum and position, (to place the turning point arbitrarily close
to the horizon). If one attributes to a particle with the rest mass $m$ a non-zero proper radius $b$, the minimal area change is for the Kerr black hole given as $\delta
A=8\pi m b$. Using this equation, one can reason as follows \cite{Bekenstein1}:
The proper radius has to be larger as the reduced Compton wavelength
of the particle, or its Schwarzschild radius, whichever is larger.
The Compton wavelength is larger for $m< 2^{-1/2}$ in Planck units
and the Schwarzschild radius is larger for $m>2^{-1/2}$ in Planck
units. For the case of reduced Compton wavelength being larger it
holds that $b>\lambda$, and the reduced Compton wavelength relates
to the rest mass as ~$m=\lambda^{-1}$. This means the following must
hold
\[m\cdot b\geq 1~~~~~~~~\rightarrow ~~~~~~~~\delta A\geq 8\pi,\]
(in Planck units). For the case of Schwarzschild radius being larger, it
holds in Planck units:
\[\delta A\geq 8\pi\cdot mb\geq 16\pi m^{2}\geq 8\pi.\]
This means one obtains
a lower bound on the Kerr black hole horizon area transition, in Planck units as
$\delta A\geq 8\pi$. As stated in the introduction, if the horizon
area spectrum is discrete and there is a lower bound on the
area quanta, the most natural spectrum to
impose is an equispaced area spectrum $8\pi\gamma\cdot n$ with
$\gamma\in O(1)$. 

What we will show is that the
same line of reasoning that Bekenstein originally used for the Kerr black hole horizon can be repeated for other spacetime horizons. By using the Bekenstein arguments we arrive at the same lower bound for the area change for all the horizons of the geometry given by the line element \eqref{line}. This suggests that also a lower bound on the area transition is a \emph{fairly general} result. The universality of the lower bound was Bekenstein's argument for an equispaced area spectrum with the quantum being approximately given by the value of the bound. One can then use the same argument to support the
suggestion by \cite{Paddy3}, to generally impose the same (and equispaced) entropy spectra
for all the spacetime horizons. Let us first explicitly calculate the area transition lower bounds for a couple of examples and then sketch a general proof for the horizons of the line element \eqref{line}. 

\subsection{deSitter spacetime}
One simple
example is the pure deSitter spacetime (with ~$f(r)=1-\Lambda r^{2}/3$).
Let us repeat the argumentation of Bekenstein: It is easy to observe that the area change is minimalized when the particle falls below the horizon radially. Therefore let us drop the
particle from its classical turning point slightly under the cosmological horizon.
$\delta$ is the radial position of the center
of mass of the particle (with the rest mass $m$), and the center of
mass is supposed to follow the geodesics of the deSitter spacetime.
The proper radius $b$ of the particle is related to $\delta$ through
the relation, ($R$ is the radial position of the deSitter
cosmological horizon):
\begin{equation}\label{int}
b=\int_{R-\delta}^{R}\sqrt{g_{rr}}~dr=\int_{R-\delta}^{R}f^{-1/2}(r)~
dr.
\end{equation}
This turns into the result:
\[\delta=\sqrt{\frac{3}{\Lambda}}\left[1-\cos\left(\sqrt{\frac{\Lambda}{3}}\cdot b\right)\right].\]
The Killing energy corresponding to a minimal area change is then:
\[E=m\cdot\sqrt{1-\frac{\Lambda}{3}\{R-\delta(b)\}^{2}}=m\cdot\sin\left(\sqrt{\frac{\Lambda}{3}}\cdot b\right).\]
Taking this relation up to the linear order in $b$ turns into:
\[E=m\cdot b \cdot\sqrt{\frac{\Lambda}{3}},\]
and in terms of horizon area transition this gives:
\[\frac{1}{4}\cdot T\cdot \delta A_{min}=\frac{1}{8\pi}\sqrt{\frac{\Lambda}{3}}\cdot\delta A_{min}=E ~~~~~~\rightarrow ~~~~~~~\delta A_{min}=8\pi\cdot m\cdot b.\]
Repeating the argumentation from the previous paragraph, one obtains precisely the same lower bound as Bekenstein for the Kerr black
hole horizon area transition!

\subsection{The inner and the outer horizons of the Reissner-Nordstr\"om spacetime}
Let us further take the maximally analytically extended
Reissner-Nordstr\"om geometry and consider both horizons (black hole
and the inner Cauchy horizon). At this stage let us skip the discussion
about the physical meaning of such geometry, (especially the discussion
about instability of the inner, Cauchy horizon). We believe that
\emph{as a matter of principle} this geometry still provides an
important and interesting example, it is not necessary to take
physically seriously the model of the black hole interior (and even
exterior) provided by the maximally analytically extended R-N
spacetime. In other words, the principal issues that we are
investigating, are believed to survive for the physically
relevant situations.

Consider dropping an \emph{uncharged} particle from above the R-N black
hole horizon, or from below the white hole inner horizon in the
maximally extended geometry. (Conveniently forget that particle
passing through the inner horizon could eventually destroy the
geometry.) Consider the upper horizon to be at the radial coordinate
$r_{+}$ and the inner horizon at the radial coordinate $r_{-}$. (The
$\pm$ signs in the notation mean that the relevant quantities relate
to the black hole outer / inner horizon case.) Then again, relating the
particle proper radius $b$ and the coordinate radius $\delta_{\pm}$
through the integral \eqref{int} one obtains (in the linear order in
$\sqrt{\delta_{\pm}}$):
\[b=\frac{2\cdot r_{\pm}}{\sqrt{r_{+}-r_{-}}}\sqrt{\delta_{\pm}}.\]
Then the Killing energy becomes (in the same order $\sqrt{\delta}$):
\[E_{\pm}=\frac{m\sqrt{\delta_{\pm}}\sqrt{r_{+}-r_{-}}}{r_{\pm}}=\frac{m b
\cdot(r_{+}-r_{-})}{2\cdot r_{\pm}^{2}}.\] The area change corresponds
to, ($\kappa_{\pm}$ are surface gravities of the two horizons):
\[\pm\frac{1}{8\pi}\cdot\kappa_{\pm}\delta
A_{(min)\pm}=\frac{1}{8\pi}\frac{r_{+}-r_{-}}{2\cdot r_{\pm}^{2}}\delta
A_{(min)\pm}=E_{\pm}~~~~\rightarrow~~~~\delta A_{(min)+}=\delta
A_{(min)-}=8\pi\cdot mb.\] 
Again, by proceeding as before this gives the same
lower bounds $\delta A_{(min)\pm}=8\pi$, for both, the black hole
and the inner Cauchy horizons.

\subsection{The Rindler horizon and a sketch of a general proof}
Let us take the Rindler line element in a suitable coordinates. It can be expressed as \cite{Paddy1}:
\begin{equation}\label{Rindler}
ds^{2}=-2\kappa x\cdot dt^{2}+(2\kappa x)^{-1}dx^{2}+dL_{\perp}^{2}.
\end{equation}
(The $\kappa$ parameter is defined to be a surface gravity of the horizon: $\kappa=[(2\kappa x)_{,x}/2]_{x=0}$.) Then let us place the particle's center of mass at a coordinate $x=\delta$ and drop it in the $x$ direction below the horizon. The following holds:
\[b=\int_{0}^{\delta}\frac{1}{\sqrt{2\kappa x}}dx=\sqrt{\frac{2}{\kappa}}\cdot\sqrt{\delta}.\]
Then the suitable Killing energy associated to the boost isometry is:
\[E=m\sqrt{2\kappa\delta}=\kappa\cdot mb,\]
and the minimum area change reads:
\begin{equation}
\frac{\kappa}{8\pi}\cdot\delta A_{min}=E~~~~~~\rightarrow~~~~~~\delta A_{min}=8\pi\cdot mb.
\end{equation}
Again, by repeating the same analysis as before we arrive at the lower bound for the horizon area change~ $\delta A_{min}=8\pi$. Note that the Rindler line element \eqref{Rindler} \emph{near the horizon} generally approximates the line element \eqref{line} , (by using the expansion $f(r)\approx f_{,r}(R)\cdot [r-R]=2\kappa\cdot [r-R]$). The proof for Rindler spacetime can therefore count as a general proof for horizons of the metric with the line element \eqref{line}.

The lower bound on the area transition turned up to be
\emph{the same} and \emph{completely independent} on the spacetime parameters, irrelevant on whether it was the inner, the outer black hole, or the
cosmological horizon. The horizon area transition bound therefore seems to be even more universal as suggested by Bekenstein: It is independent both on the spacetime parameters and the particular horizon. If we
accept (for each of the horizons) that the horizon area semi-classical spectrum is equispaced,
it is natural to relate it to the lower bounds in Planck units as:
\begin{equation}\label{spectrum}
A_{i n}= 8\pi\gamma_{i}\cdot n,~~~~~~~~~~\gamma_{i}\in O(1),
\end{equation}
 where $i$
labels the different horizons. It seems to be also natural to claim that
$\gamma_{i}$ is the same for all the horizons. The entropy
quantization suggested by \cite{Paddy3} gives (for general relativity) precisely the spectrum
\eqref{spectrum} with the universal choice $\gamma_{i}=1$,~ which certainly
fits the bill.

\section{Maggiore's conjecture and the information from the asymptotic quasinormal modes}

\subsection{Some reasons supporting the conjectured connection between the spacetime thermodynamics and the asymptotic quasi-normal modes \label{discussionMaggiore}}

Let us discuss some basic rasons why to suspect a connection between the asymptotic quasi-normal modes and the thermodynamics of horizons. We will discuss it in a more generalized setting as is done in the original papers \cite{Hod, Maggiore}, where the discussion is related to the Schwarzschild spacetime.

The dynamics of fields near horizons of the line element \eqref{line} can be generally approximated by a free field dynamics. This can be trivially shown for the simplest example of a massless scalar field, where the dynamics is given by:
\begin{equation}\label{freescalar}
\frac{1}{\sqrt{-g}}\partial_{\mu}(g^{\mu\nu}\sqrt{-g}\partial_{\nu}\Psi)=0. 
\end{equation}
By decomposing the scalar field throgh multipoles $Y_{\ell m}$ as:
\[\Psi=\sum_{\ell, m}\frac{\psi_{\ell m}}{r}Y_{\ell m},\]
and turning into the tortoise coordinate
\[x=\int^{r}\frac{dr'}{f(r')},\]
the equation \eqref{freescalar} turns into:
\begin{equation}\label{freescalar2}
-\frac{\partial^{2}\psi_{\ell m}}{\partial t^{2}}+\frac{\partial^{2}\psi_{\ell m}}{\partial x^{2}}-V(\ell,x)\psi_{\ell m}(x)=0,
\end{equation}
with the potential
\[V\{\ell,r(x)\}=f(r)\left[\frac{\ell(\ell+1)}{r^{2}}+\frac{2 f_{,r}(r)}{r}\right].\]
We see that near the horizons where from the definition $f(r)\approx 0$, the potential becomes $V(r)\approx 0$ and the field behaves as a free field. The perturbation equations for more complicated perturbative fields is more difficult to derive, but eventually all of the fields can be reduced to degrees of freedom following the dynamics given by the equation \eqref{freescalar2}, with some suitable potential $V(x)$. Moreover, the potential is in every case vanishing as one approaches the horizon. (Therefore the scalar field is as in many situations a good simple model for physically much more realistic fields.) 

Now let us consider that with some approximation, (the quasi-normal modes usually do not form a complete system \cite{Nollert}), one can
decompose the field into discrete quasi-normal modes. Considering this, the situation near the horizon resembles a classical emission by a system that can oscillate around some configuration only with certain discrete frequencies (such as a string). Furthermore the oscillations decay due to the radiation of energy. Let us take this picture seriously: it allows us to associate with the horizon a set of discrete frequencies related to the quasi-normal modes of the field. 

Now let us follow the Bohr correspondence principle: ``In the limit of high quantum numbers the transition frequencies are equal to the classical oscillation frequencies''. What was said in the previous paragraph suggests that one can attempt to identify the horizon transition frequency in the semi-classical limit with the quasi-normal modes. But we have a discrete infinity of quasi-normal modes, so which quasi-normal modes shall we pick? There is no information about the underlying quantum theory that could provide guidance, but there is one basic observation from the known physics: the transition between the systems quantum levels in the semi-classical limit shall have relaxation times close to zero. 
Therefore it is reasonable to look for the ``correct'' frequencies in the limit of the high damping. Remarkably enough, the behavior of the frequencies in this limit is independent on the wave mode number $\ell$. Hod \cite{Hod} suggested to associate the classical oscillation frequency simply with the real part of the quasinormal modes, as \[\lim_{n\to\infty} \omega_{R n}\doteq\omega_{R\infty}.\] In particular Hod considered quasi-normal modes of gravitational perturbations and since for such perturbations holds:
\[\lim_{n\to\infty} \omega_{R n}=T \ln(3),\]
Hod \cite{Hod} obtained the Bekenstein entropy quantization of the ``statistical'' form \cite{BekensteinMukhanov}:
 \[S=\ln(3) n.\]

However, over the period of time certain arguments were raised against Hod's reasoning \cite{Maggiore, Zhidenko}. One very important objection is that the $\omega_{R\infty}$ quantity is not universal even for the Schwarzschild spacetime, but depends on the spin of the perturbation. Even worse, when turning to a more complicated spacetime, like Reissner-Nordstr\"om, Schwarzschild-deSitter, Reissner-Nordstr\"om-deSitter, or to non-spherically symmetric ones (like Kerr, Kerr-Newman, Kerr-Newman-deSitter), there is no indication that the quantity $\omega_{R\infty}$ is reasonably defined even for a specific field. (The formulas for the quasi-normal frequencies for more general spherically symmetric spacetimes are derived in \cite{Motl, Andersson, Natario1, Natario2, Shanki2}. These formulas seem to suggest that the $\omega_{R\infty}=T\ln(k),~ k\in\mathbb{N_{+}}$~ result is in fact a lucky coincidence of the fact the spacetime has a single horizon.)

However Maggiore \cite{Maggiore} made a very interesting observation, such that can be shown to resolve all the main objections against Hod's conjecture: Since the quasi-normal oscillations are damped oscillations, it is not reasonable to identify the classical oscillation frequencies with the real part of the quasi-normal frequency, but rather with the proper frequency of the damped oscillator, given as 
\begin{equation}\label{proper}
\omega_{0n}=\sqrt{\omega_{Rn}^{2}+\omega_{In}^{2}}.
\end{equation}   
Let us further give our account on what is happening: One can preserve the idea that the only relevant perturbations are those that represent transitions with very low relaxation times, and these are the perturbations composed (with some approximation) only from highly damped quasinormal modes. Then via Bohr's principle it is natural to see the sequence of frequencies \eqref{proper} to describe the transition frequencies from some background state with quantum number $N$ to a state with $N+n$. The frequency corresponding to a transition from $n\to n-1$ is then naturally:
\begin{equation}\label{proper2}
\Delta_{(n,n-1)}\omega_{0n}.
\end{equation}

Since we deal with the highly damped frequencies it is reasonable to take the expression \eqref{proper2} with a limit $n\to\infty$. This limit should give sensible answer even if one considers that the transitions between $N$ and $N+n$ are upper bounded by the fact that we are considering only ``small'' perturbations around background spacetime with some particular values of parameters. (Indeed one can see from the numerical results that \eqref{proper2} is very well approximated by its value at $n\to\infty$ long time before one would expect that the transition is outside the scope of the linearized theory.)
However, it should be stressed that at no point one can be self-confident enough to claim that this is more than a suggesive way of thinking. The perturbation cannot be decomposed in quasi-normal modes exactly and many subtle points in the logic could become potentially fatal. This is the reason why the upper statement, despite being useful and interesting, is still only a \emph{conjecture}. On the other hand as we will see the conjecture is very fruitful and has a potential to explain a great universality in the results obained over the years.

There is another evidence that supports the connection between the
quasi-normal modes and the horizon thermodynamics as suggested by
Maggiore. It can be reasonably generally observed that the structure
of the spacing in the imaginary part of the asymptotic frequencies
depends only on the tails of the scattering potential, in other words only on
the characteristics of the horizons. Adding a matter field between
the horizons, such that modifies the geometry
between the horizons, but keeps the structure of the horizons unchanged,
does not affect the structure of the imaginary part of the
asymptotic frequencies! (Of course the field whose quasi-normal
frequencies we are computing must interact with the field that was added
only via gravity.) (All this can be explicitly seen in the results derived in the papers \cite{Shanki1, Shanki2, Medved} and one does not have to even worry about the general relativity theory.)

As another indirect support for the conjecture one can see the results from the paper written by me and my colaborators \cite{Skakala3}. (See also \cite{Skakala4}.) We have tried to answer the question of what happens with the asymptotic frequencies, if a generic spherically symmetric static asymptotically flat spacetime does not have an (absolute) horizon. The answer we found was that in such case there will be \emph{no} asymptotic quasi-normal modes. (For a fixed $\ell$ there will exist a bound on the frequencies.) This means there seems to be some type of general connection between the existence and properties of the asymptotic quasi-normal modes and the existence and properties of (absolute) horizons, as could be assumed from the Maggiore's (or Hod's) conjecture. (Note that the non-existence of the asymptotic frequencies was shown assuming spherical symmetry and staticity, but if one believes that the zero angular momentum limit is for the most fundamental characteristics of the frequencies non-singular, then one can see it as an evidence that such a result survives for a more general, axially symmetric spacetimes.)

Hod's/Maggiore's conjecture in fact suggests that the classical general
relativity matches the classical limit of some underlying quantum
theory in a more ``sensitive'' way, than is usual in the standard
examples of classical theories. The classical description becomes better
and better approximation as the black hole grows, and the energy
quanta emitted by the black hole go to zero more and more
approximately as $\sim T$. Instead of approximating this picture by
a continuum, (assuming that Maggiore's conjecture is being correct),
general relativity matches the sequence of quanta already before
reaching zero. Therefore the information that is carried by the general relativity
about thermodynamics is in such case deeper than expected, since general relativity would carry some
information about the discrete structure that appears under
the continuous spacetime.

\subsection{The behavior of the asymptotic frequencies for the spherically symmetric multi-horizon spacetimes}

Here we will consider three spherically symmetric spacetimes:
Reissner-Nordstr\"om-deSitter (R-N-dS), Schwarzschild-deSitter
(S-dS) and Reissner-Nordstr\"om (R-N) spacetime. One might wonder, why we are
considering these three spacetimes separately and not the latter
ones as a special case of the R-N-dS spacetime. The reason is that
the zero limits for $M,Q,\Lambda$ parameters are for the quasi-normal frequencies
sometimes singular. (There are also reasons to believe that any
Maggiore-like interpretation for the frequencies could be
in this limit singular, as we will argue later.) The transcendental
formulas for the asymptotic quasi-normal frequencies for all the
three, R-N, S-dS, R-N-dS spacetimes can be written (for uncharged scalar, gravitational, electromagnetic fields) as \cite{Andersson, Natario1, Natario2}:
\begin{equation}\label{frequencies0}
\sum_{A=1}^{K}Z_{A}\exp\left(\sum_{i=1}^{N}B_{Ai}\cdot\frac{\omega}{T_{i}}\right)=0.
\end{equation}
Here $Z_{A}, B_{Ai}$ are $K\times 1$ and $K\times N$ matrices composed of integers, $N$ is the
number of horizons in the spacetime ($N=2,3$ in our cases), $K$ is some integer that depends on the particular formula and
$T_{i}$ are the temperatures of the horizons. (Note that we
consider here also the inner, Cauchy horizon, assuming the horizon
temperature from the periodicity of the Euclidean solution, hence
$T=|\kappa|/2\pi$. Here $\kappa$ is the surface gravity of the
horizon.) 

The formulas \eqref{frequencies0} were known for some time (since \cite{Andersson, Natario1, Natario2}), but it is known
that they do not have analytic solutions (unlike simpler
Schwarzschild case). On the other hand, despite of the fact that they
do not have analytic solutions, we have shown \cite{Visser3, SkakalaPhD} that a lot of
\emph{important} analytic information can be extracted about the
behavior of the solutions. (Important from the point of view of
Maggiore's conjecture.) 

We have shown \cite{Visser3} that the solutions for all the three cases (for
a field with spin $s$) follow a simple pattern: If the ratio of
all the horizon temperatures is a \emph{rational} number, the frequencies
split into a \emph{finite} number of families labeled by $a$, given
as:
\begin{equation}\label{frequencies}
\omega_{a s n}=\hbox{(offset)}_{a s}+i n\cdot 2\pi\cdot
\hbox{lcm}(T_{1},T_{2},...,T_{N})+O(n^{-1/2}).
\end{equation}
Here
$lcm$ means ``the least common multiple'' of the temperatures, hence\footnote{The formulas \eqref{frequencies0} derived in \cite{Andersson, Natario1, Natario2} are generally accepted (and numerically confirmed) as the exact formulas for the asymptotic sequence of the highly damped frequencies. There are also known results for the asymptotic quasi-normal frequencies of the S-dS spacetime obtained via Born approximation \cite{Medved, Paddy6}, or approximation by analytically solvable potentials \cite{Visser1, Visser2}. Although on the first sight these results might seem to be inconsistent with the behavior described by \eqref{frequencies0}, one has to realise that both, the methods of Born approximation and of analytically solvable potentials, allow for each frequency an error of the order of unity. Considering this, the results obtained by Born approximation and by analytically solvable potentials can, (at least in some cases), describe some form of averaged behavior of the solutions of \eqref{frequencies0} and hence be fully consistent with the results of \cite{Medved, Paddy6, Visser1, Visser2}.}: \[
\hbox{lcm}(T_{1},T_{2},...,T_{N})=p_{1}T_{1}=p_{2}T_{2}=...=p_{N}T_{N},
\] where $p_{1},...,p_{N}$ are relatively prime integers.
It can be also proven that: \emph{In case ratio of two of the
horizon temperatures is irrational, there is no infinite
periodic subsequence of the sequence of asymptotic quasi-normal
frequencies.}

This behavior shows almost the same universality as in the Schwarzschild
case and is fascinating: How can the frequencies depend (in case of
R-N, R-N-dS spacetimes) also on the inner Cauchy horizon, which means on the horizon
that is in the causally disconnected region? The rather formal sounding explanation
is that the frequencies must depend on all the $\Lambda, M,
Q$ parameters and one can always transform the $\Lambda, M, Q$
variables into the $T_{1}, T_{2}, T_{3}$ variables. (Or for R-N
spacetime the $M,Q$ variables to $T_{1}, T_{2}$ variables.) This
means the frequencies must depend also on the temperature of the
Cauchy horizon. But why does one observe such a simple and universal
structure, which is in fact symmetric under the
exchange of\footnote{This is again a general feature of static spherically symmetric two horizon spacetimes\cite{Shanki2, Medved}, independent on general relativity. On the other hand Maggiore's conjecture depends on the general relativity.} ~$T_{i}\leftrightarrow T_{j}$~?  Note
that this cannot be explained by the unit analysis. A general unit
analysis just tells you that the gap in the spacing of the imaginary
part of the families is (for R-N-dS, for example) given as:
\[T_{1}\cdot F\left(\frac{T_{1}}{T_{2}}, \frac{T_{2}}{T_{3}},n,s,\ell\right),\]
where $F$ is an arbitrary function. (Let us mention that \cite{Castro} offers some answers to the question of why there is the observed dependence of the asymptotic frequencies on the temperature of the Cauchy horizon. The authors of \cite{Castro} use the Kerr-CFT correspondence, and the conclusions seem to be consistent with the conclusions we make in this paper.)

Consider (in the imaginary part) monotonically ordered union of the families of the asymptotic frequencies. In \cite{Skakala2} we have shown, that it is reasonable to expect the limit in the
spacing of the imaginary part of such a sequence \emph{not} to exist. The limit can be shown
not to exist also in the case the ratio of some of the temperatures is
an irrational number. But despite of this unpleasant property, there is a way of how to interpret
the observed behavior of the frequencies along the lines suggested
by Maggiore.

\subsection{The interpretation of the frequencies for the Schwarzschild-deSitter (S-dS) spacetime}

Let us first \cite{Skakala3} consider field propagating in the maximally
analytically extended S-dS spacetime. The field does not change the cosmological constant and
propagates through the white hole horizon and penetrates eventually
the black hole horizon. The first law of thermodynamics
\eqref{firstlaw} turns into:
\begin{equation}\label{firstlaw2}
T_{BH}\delta S_{BH}+T_{CH}\delta S_{CH}=0.
\end{equation}
(We are labeling the two horizons in the same way as before.) If both the
horizons have equispaced entropy spectra of the form $S_{i
,n_{i}}=2\pi\gamma\cdot n_{i}$, then the first law of thermodynamics
gives:
\[\frac{T_{CH}}{T_{BH}}=-\frac{\Delta S_{BH}}{\Delta
S_{CH}}=-\frac{n_{1}}{n_{2}},\] which means the temperatures of
the two horizons must have rational ratio. As we have
seen, the quasi-normal modes represent the quanta of Killing energy, and from the equation \eqref{areavar} we see that
the energy which flowed through the horizons (sufficiently slowly, as we are in thermodynamics) corresponds to $\delta M$. Therefore by applying Maggiore's conjecture one obtains:

\begin{eqnarray}\label{frequencies}
\delta M=T_{BH}\delta
S_{BH}=-T_{CH}\delta S_{CH}=T_{BH}\cdot m_{BH}\cdot 2\pi\gamma=~~~~~~~~~~~~~~~~~~~~~~~~~~~~\nonumber\\
=T_{CH}\cdot m_{CH}\cdot
2\pi\gamma=\lim_{n\to\infty}\Delta_{(n,n-1)}\omega_{nI}.~~~~~~~~
\end{eqnarray}

Here $m_{BH}, m_{CH}$ are some integers and express the fact that the
entropies can change only by the integer multiples of $2\pi\gamma$.
Considering the fact that $\delta M$ has to be a smallest possible
quantum, such that in the same time fulfills the equation \eqref{frequencies}, one obtains:

\begin{equation}
\delta M=2\pi\gamma\cdot\hbox{lcm}(T_{BH},
T_{CH})=\lim_{n\to\infty}\Delta_{(n,n-1)}\omega_{nI}.
\end{equation}
This is precisely what we observe in the spacing in each of the
finite number of families, if we fix $\gamma=1$ ! It means, that only the equispaced families of modes should carry the physical information about the mass quanta emitted in the horizon state transition\footnote{The fact that this particular information is encoded in the families of frequencies, rather than the asymptotic sequence itself, is a necessary consequence of the fact that the frequences are continuous functions of the spacetime parameters.}. This is what we consider to be a generalization of Maggiore's original hypothesis, which was used for the Schwarzschild spacetime.

\subsection{The interpretation of the frequencies for the Reissner-Nordstr\"om (R-N) and the Reissner-Nordstr\"om-deSitter (R-N-dS) spacetime}

Let us further consider maximally analytically extended R-N and R-N-dS
spacetimes. Furthermore, let us consider that there is an \emph{uncharged} field (also $\Lambda$ stays fixed) flowing from the white hole horizon towards, and through the cosmological horizon of the R-N-dS spacetime, (or towards the asymptotic infinity of the R-N spacetime). (Alternatively let us consider the Hawking radiation coming from the black hole horizon.) We claim again, that these considerations have meaning \emph{as a matter of principle}, despite of the fact that physically the models are unrealistic for various different reasons: a) there are no charged astrophysical macroscopic objects, so one cannot consider the charged geometry seriously\footnote{However, the (interior) Reissner-Nordstr\"om geometry is often considered as a good qualitative approximation for the Kerr geometry. The external Kerr geometry is, of course, the one that is supposed to be physically relevant.}, b) even if one considered astrophysical charges, the white hole cannot appear in a solution arising from a stellar colapse, c) the collapse scenario is supposed to lead to a mass inflation singularity replacing the inner horizon of the R-N (or R-N-dS) geometry\footnote{On the other hand let us mention here the following two points: 1. There is a regime in the R-N-dS spacetime in which the Cauchy horizon is stable \cite{Poisson}. 2. The mass inflation singularity is only a weak singularity \cite{Hod2, Hod3}. The tidal forces experienced by an infalling observer are finite and it can be suggested that an infalling observer could still cross the singularity\cite{Hod2, Hod3}. This would mean it can still act as some kind of horizon.}.  Note that the instability of Cauchy horizons is a lucky consequence for those who want general relativity to be predictable; in that case it is necessary to exclude the possibility that timelike singularities could form. (Or in other words the strong cosmic censor conjecture holds.) 

However let us consider the previous example and take seriously (at least on some limited time scale) also the inner Cauchy horizon. Consider explicitly the R-N-dS spacetime. (It is straightforward to modify this considerations for the R-N spacetime, so we won't explicitly do it). Then the equation \eqref{firstlaw2} still holds and from \eqref{areavar} we see that the energy flowing through the horizons corresponds again to $\delta M$. But, since $\delta\Lambda=\delta Q=0$ the variation in $\delta M$ has to fulfill also:
\[\delta M=-T_{IH}\delta S_{IH}.\] 
Here $T_{IH}, S_{IH}$ are the entropy and temperature of the inner, Cauchy horizon. (For the other two horizons we keep the notation from the previous section.) This means, that by applying the same logic as in the previous section, (and assuming that all the horizons have the same, discrete and Bekenstein-type spectra), one comes to the following conclusion: If one allows processes, such that change only the $M$ parameter, all the temperatures must have \emph{rational} ratios\footnote{For the R-N spacetime the rational ratios follow directly from the fact that: $T_{IH}S_{IH}=T_{BH}S_{BH}$.}. Further, by using the same logic as before, one obtains:
\begin{eqnarray}\label{frequencies2}
\delta M=-T_{IH}\delta S_{IH}=T_{BH}\delta
S_{BH}=-T_{CH}\delta S_{CH}=~~~~~~~~~~~~~~~~~~~~~~~~~~~~~~~~~~~~~~~~~~\nonumber\\
=T_{IH}\cdot m_{IH}\cdot 2\pi\gamma=T_{BH}\cdot m_{BH}\cdot 2\pi\gamma=T_{CH}\cdot m_{CH}\cdot
2\pi\gamma=\lim_{n\to\infty}\Delta_{(n,n-1)}\omega_{nI}.~~~~~~~~~
\end{eqnarray}
($m_{IH}, m_{BH}, m_{CH}\in\mathbb{Z}$.) Consider that we are again interested in the smallest allowed quantum, such that fulfills the equation \eqref{frequencies2}. Then one obtains again:
\begin{equation}\label{R-N-dS}
\delta M=2\pi\gamma\cdot\hbox{lcm}(T_{IH}, T_{BH},
T_{CH})=\lim_{n\to\infty}\Delta_{(n,n-1)}\omega_{nI}.
\end{equation}
This is again the observed behavior for the families of the asymptotic frequencies, such that one (again) fixes the $\gamma$ parameter in the usual way: $\gamma=1$. (As in case of S-dS spacetime, this provides the same generalization of Maggiore's conjecture for the multi-horizon spacetimes.) (Let us add here that the same result, the universal entropy (area) spectra given as $2\pi\cdot n$ ($8\pi\cdot n$), was obtained for the multi-horizon spacetimes in \cite{Kwon}. The authors used different argumentation as we did here, but their reasoning was also partly related to the asymptotic quasi-normal modes.) 

\subsection{Discussion}

Let us briefly discuss the condition for the rational ratios of the horizon temperatures. It is used as an important middle step in the argumentation, connecting the behavior of the frequencies with the quantization of entropy for the different horizons. But one in fact does not have to mention the rational ratios at all, and use directly the formula \eqref{frequencies2} leading to the equation \eqref{R-N-dS}, where the temperatures rational ratio follows implicitly from the result. (``The least common multiple'' must be a well defined function for the temperatures.) Since irrational numbers are dense in the set of rational numbers, it would seem as a miracle, if the rational ratios would say something fundamental about the nature. (Something like the general theory of relativity carrying information about some quantization of temperatures.) Let us keep this question open, but let us mention that similar rational ratio condition for the temperatures of the two horizons in the S-dS spacetime had been derived in \cite{Paddy7}. In particular, the rational ratios follow from the condition of existence of a global thermodynamical equilibrium in the S-dS spacetime.

Let us also mention that although the gap in the spacing between the frequencies is a continuous function of the $\Lambda, Q, M$ parameters, our interpretation of the frequencies is singular as $Q\to 0$, or $\Lambda\to 0$ ! Is this a problem? I would like to argue that certainly not. We are constrained to the semi-classical regime and the interpretation is dependent on us taking the properties of all the horizons seriously. By taking such limits as $Q\to 0$, $\Lambda\to 0$, one of the horizons becomes increasingly small and its entropy spectrum eventually deviates from the semi-classical regime. Hence it has to be expected that the thermodynamical interpretation for the frequencies is for the limits $Q\to 0$, or $\Lambda\to 0$ singular.

As a last point let us discuss, whether our suggestion that areas of all the different horizons are semi-classically equispaced can appear troublesome from the point of view of some known physics. Is there some basic constraint by the well known physics that can interfere with the area quantization rules? Such constraint might be the quantization of charge. Any black hole's charge should be given as a multiple of the elementary charge, hence $Q=N\cdot e$. For the Reissner-Nordstr\"om black hole the relation between areas of the horizons and the charge is simple: \[A_{1}A_{2}=K\cdot Q^{4}=(k_{\epsilon}^{2}G^{2}/c^{8})\cdot Q^{4}.\] (The value of $K$ is approximately of the order $\sim 10^{-68} m^{4}/C^{4}$.)
This means
\begin{equation}\label{charge}
n_{1}n_{2}=\frac{1}{(8\pi l_{p}^{2})^{2}}K e^{4}N^{4}.
\end{equation}
The dimensionless constant 
\[\frac{1}{(8\pi l_{p}^{2})^{2}}K e^{4}\sim 10^{-6}.\]
Take the weakest possible interpretation of what the equispaced spectra of the two horizons mean and that is: The areas of the horizons can be given only as $A_{i}=8\pi n_{i}$ and for each relevant couple of natural numbers $n_{1}, n_{2}$ there is a quantum state in which the area values are obtained. (The ``relevance'' of the numbers is constrained by the condition that both $n_{1}, n_{2}>>1$ and correspond to a black hole far from extremality.)
But then the condition \eqref{charge} hardly represents any problem as it just tells that for arbitrary $N$ you have to find with a ``good enough'' approximation some natural numbers $n_{1}, n_{2}$, such that fulfil the upper condition. In fact the right side of the equation is always integer with the error $\leq 1/2$ and for large $n_{1}\cdot n_{2}$ this always gives good approximation. (Note that an entirely different thing would be to impose, that the quantum states can be labeled \emph{independently} by two different quantum numbers labeling the area spectra of the two horizons. This is a \emph{stronger} condition as we impose, and trivially contradicts our results\footnote{We agree that this would be closer to the Bekenstein proposal from \cite{Bekenstein3}.}, considering that electromagnetic charge has to be quantized in the physically relevant way.)   

\section{Conclusions}

As we already discussed our results in several places in the paper, let us briefly conclude: The key statement of this review of our work is to extend the idea that black hole horizon has entropy with a semi-classically equispaced spectrum to much more general spacetime horizons. Such a generalization was already proposed by using different arguments in \cite{Paddy3}. We support the idea of \cite{Paddy3} by adding two basic arguments: ~1. we generalize Bekenstein's lower bound on the horizon area transition to much more general horizons than only the black hole horizon, ~2. we obtain the same suggestion for the spectra via the information provided by the asymptotic quasi-normal frequencies of the Reissner-Nordstr\"om, Schwarzschild-deSitter and Reissner-Nordstr\"om-deSitter spacetimes, (in the way proposed by Maggiore).

\bigskip

~\\
{\bf Acknowledgments:} This research was supported by FAPESP.

\end{document}